\documentclass[conference,table,xcdraw,10pt,final]{IEEEtran}

\usepackage{microtype}
\usepackage{flushend}

\usepackage[utf8]{inputenc}

\usepackage[T1]{fontenc}
\usepackage{amsmath}
\usepackage{bm}
\usepackage{xspace}
\usepackage{booktabs}
\usepackage[binary-units=true]{siunitx}
\usepackage{bigfoot}

\usepackage{hyperref}

\usepackage{todonotes}

\usepackage{siunitx}

\usepackage{subfig}

\usepackage{dblfloatfix}

\usepackage{multirow}
\usepackage{pifont}

\usepackage{textcomp}

\usepackage{amsfonts}

\usepackage[absolute]{textpos}
\textblockorigin{1.7cm}{26cm}

\title{FIRESTARTER~2: Dynamic~Code~Generation~for~Processor~Stress~Tests}

\author{
	\IEEEauthorblockN{Robert Sch\"{o}ne¹, Markus Schmidl², Mario Bielert³, Daniel Hackenberg³}
	\IEEEauthorblockA{
		Center for Information Services and High Performance Computing (ZIH)\\
		Technische Universit\"{a}t Dresden, 01062 Dresden, Germany \\
		¹robert.schoene@tu-dresden.de, ²markus.schmidl@mailbox.tu-dresden.de, ³\{firstname.lastname\}@tu-dresden.de
	}
}

\begin{document}

	\newcommand{\mycite}[1]{\textit{``#1''}}

	\newcommand{\figref}[1]{Figure~\ref{fig:#1}}
	\newcommand{\tabref}[1]{Table~\ref{tab:#1}}
	\newcommand{\secref}[1]{Section~\ref{sec:#1}}
	\newcommand{\lstref}[1]{Listing~\ref{lst:#1}}
		
	\maketitle
	
	\begin{abstract}
		Processor stress tests target to maximize processor power consumption by executing highly demanding workloads.
		They are typically used to test the cooling and electrical infrastructure of compute nodes or larger systems in labs or data centers.
		While multiple of these tools already exists, they have to be re-evaluated and updated regularly to match the developments in computer architecture.
		This paper presents the first major update of FIRESTARTER, an Open Source tool specifically designed to create near-peak power consumption.
		The main new features concern the online generation of workloads and automatic self-tuning for specific hardware configurations.
		We further apply these new features on an AMD Rome system and demonstrate the optimization process. 
		Our analysis shows how accesses to the different levels of the memory hierarchy contribute to the overall power consumption.
		Finally, we demonstrate how the auto-tuning algorithm can cope with different processor configurations and how these influence the effectiveness of the created workload.
	\end{abstract}


\begin{textblock}{1}(0,0)
\textbf{© 2021 IEEE under \href{https://doi.org/10.1109/Cluster48925.2021.00084}{DOI 10.1109/Cluster48925.2021.00084}.} Personal use of this material is permitted. Permission from IEEE must be obtained for all other uses, in any current or future media, including reprinting/republishing this material for advertising or promotional purposes, creating new collective works, for resale or redistribution to servers or lists, or reuse of any copyrighted component of this work in other works.
\end{textblock}
	
	\section{Introduction}
	\label{sec:intro}
\begin{figure}[t]
		\centering
		\includegraphics[width=.8\columnwidth]{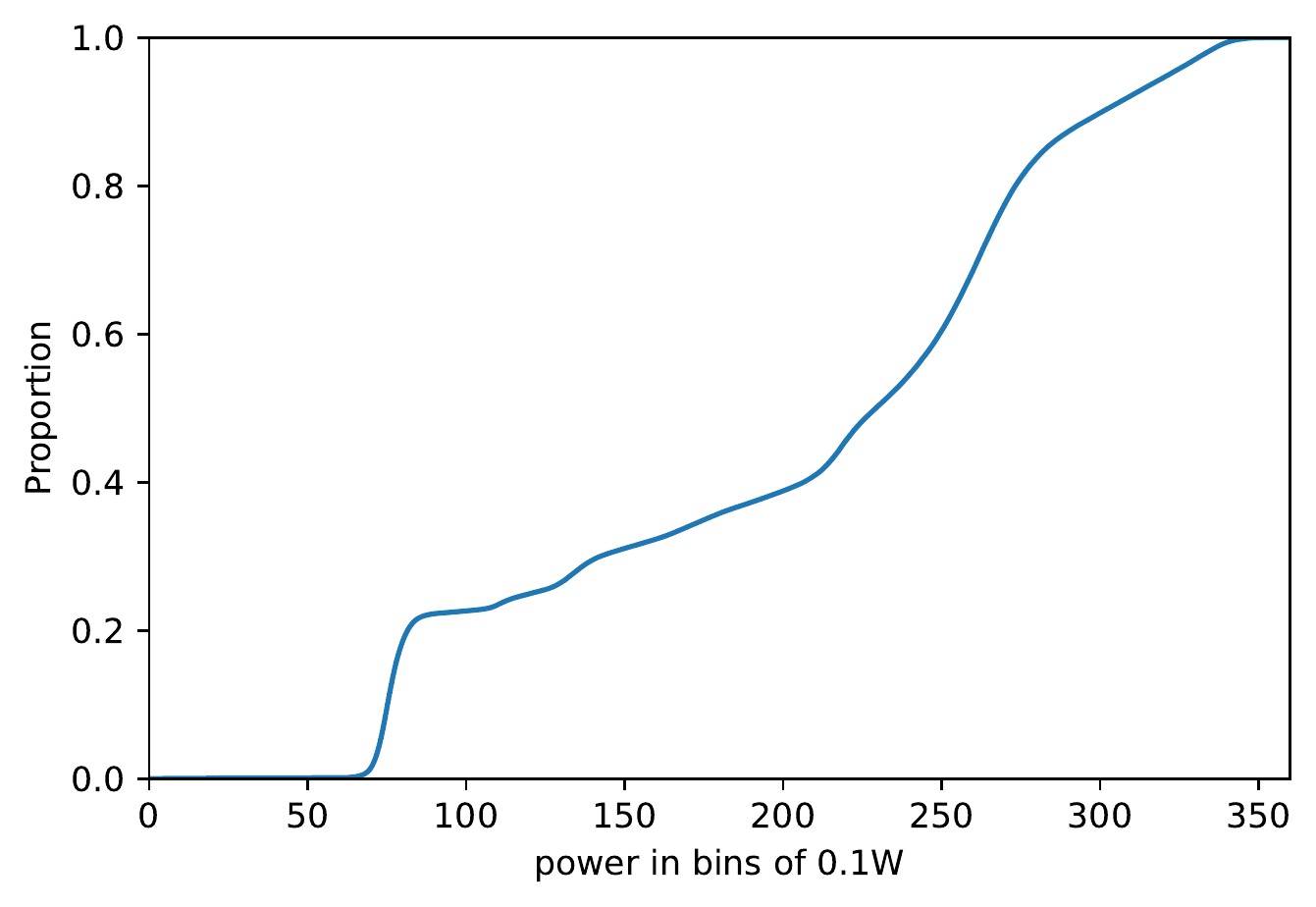}
		\caption{\label{fig:taurus6000power}Cumulative distribution of power consumption for 612 Haswell nodes of the taurus HPC system at TU Dresden in 2018. All datapoints (\SI{1}{Sa \per \second} per node) are aggregated (mean of \SI{60}{\second}) and binned into \SI{0.1}{\watt} bins.}
	\end{figure}

	The power consumption of computer systems and their components is highly varying and depending on the executed workloads.
	Most of the time, the available power budget is not used to the full extend.
	\figref{taurus6000power} exemplifies this by depicting a cumulative distribution of the power consumption of 612 Haswell nodes of the Taurus HPC system at TU Dresden over one year.
	While it shows that some workloads use up to \SI{359.9}{\watt}, others reach much less with a steep incline between \SI{50}{\watt} and \SI{100}{\watt} created by idle power consumption.
	However, the power supply and cooling infrastructure must be designed to function under the worst possible conditions.
	
	Processor and computer stress tests support system integrators and administrators to design and configure systems for such circumstances.
	By stressing specific components to the greatest extent, which results in high power consumption and thermal output, they help to find potential problems, including but not limited to inducing bit-flips~\cite{bit-flips}, processor clock throttling~\cite[Section 11.1.5.1]{acpi}, and system shutdowns~\cite[Section 11.1.6]{acpi}.
	However, with changes in the processor architecture, stress tests have to be re-evaluated and updated regularly.
	In this paper, we present recent developments in FIRESTARTER, which enable users to find higher demanding working points for their specific systems.
	The paper is structured as follows:
	We describe how processor power consumption can be maximized and analyze some of the most popular stress tests in \secref{background}.
	In \secref{firestarter2}, we analyze and optimize FIRESTARTER, a stress test to maximize power consumption and thermal output of x86\_64 processors, DRAM, and NVIDIA GPUs.
	We demonstrate the performance of our new approach in \secref{zen2} and conclude the paper with a summary and an outlook in \secref{summary}.
	
	\begin{table*}[b]
		\centering
		\caption{\label{tab:overview}Overview over different stress tests for Linux}
		\begin{tabular}{lrccccccc}
			\rowcolor[HTML]{C0C0C0}
			&
			& \multicolumn{4}{c}{\cellcolor[HTML]{C0C0C0}Stressed Components} &   Error        &       Define new         &  Compiler(-flag) \\
			\rowcolor[HTML]{C0C0C0}
			\multirow{-2}{*}{\cellcolor[HTML]{C0C0C0}Benchmark} &
			\multirow{-2}{*}{\cellcolor[HTML]{C0C0C0}Workload} & Processor       & Memory      & GPU              & Network      & check & algorithms &  independent \\ \hline
			FIRESTARTER 1~\cite{Hackenberg_2013_FIRESTARTER}  & artificial workloads & \ding{51}\ding{51}\ding{51} & \ding{51}\ding{51}\ding{51} & \ding{51}      &   \ding{55}           &   \ding{55} & \ding{51} (template) &  \ding{51}  \\
			\rowcolor[HTML]{EFEFEF}
			Prime95~\cite{mprime_website}     & Mersenne prime hunting    & \ding{51}\ding{51}  & \ding{51}\ding{51}\ding{51}   &   \ding{55}               &  \ding{55}            &  \ding{51} &   \ding{55} &  \ding{51}  \\
			Linpack~\cite{linpack}    & linear algebra   & \ding{51}\ding{51}   & \ding{51}\ding{51} & \ding{51}      &  \ding{51}¹   &  \ding{51} &   \ding{55} &  \ding{51}³  \\
			\rowcolor[HTML]{EFEFEF}
			stress-ng~\cite{stressng_website}  & various (e.g., search, sort) & \ding{51}  & \ding{51}   &  \ding{55} & \ding{55}  &  \ding{51}$^4$ &   \ding{51} (source code) &   \ding{55}     \\
			eeMark~\cite{Molka_12_eeMark}   & artificial workloads    & \ding{51}\ding{51}   & \ding{51}\ding{51} & \ding{55}      &  \ding{51}   &  \ding{51}² &   \ding{51} (template) &  \ding{55}    \\
			\rowcolor[HTML]{EFEFEF}
			FIRESTARTER~2   & artificial workloads & \ding{51}\ding{51}\ding{51} & \ding{51}\ding{51}\ding{51} & \ding{51}      &   \ding{55}           &   \ding{51} & \ding{51} (runtime) &  \ding{51} \\ \hline
			\multicolumn{9}{l}{¹ via MPI in High Performance Linpack (HPL) $|$ ² no check for bit-flips $|$ ³ but library-dependent (BLAS/LAPACK) $|$ $^4$ Only for some workloads}\\
		\end{tabular}
	\end{table*}
	
	\section{Background and Related Work}
	\label{sec:background}
	\subsection{Processor Power Consumption}
\begin{figure}[t]
		\centering
		\includegraphics[width=\columnwidth]{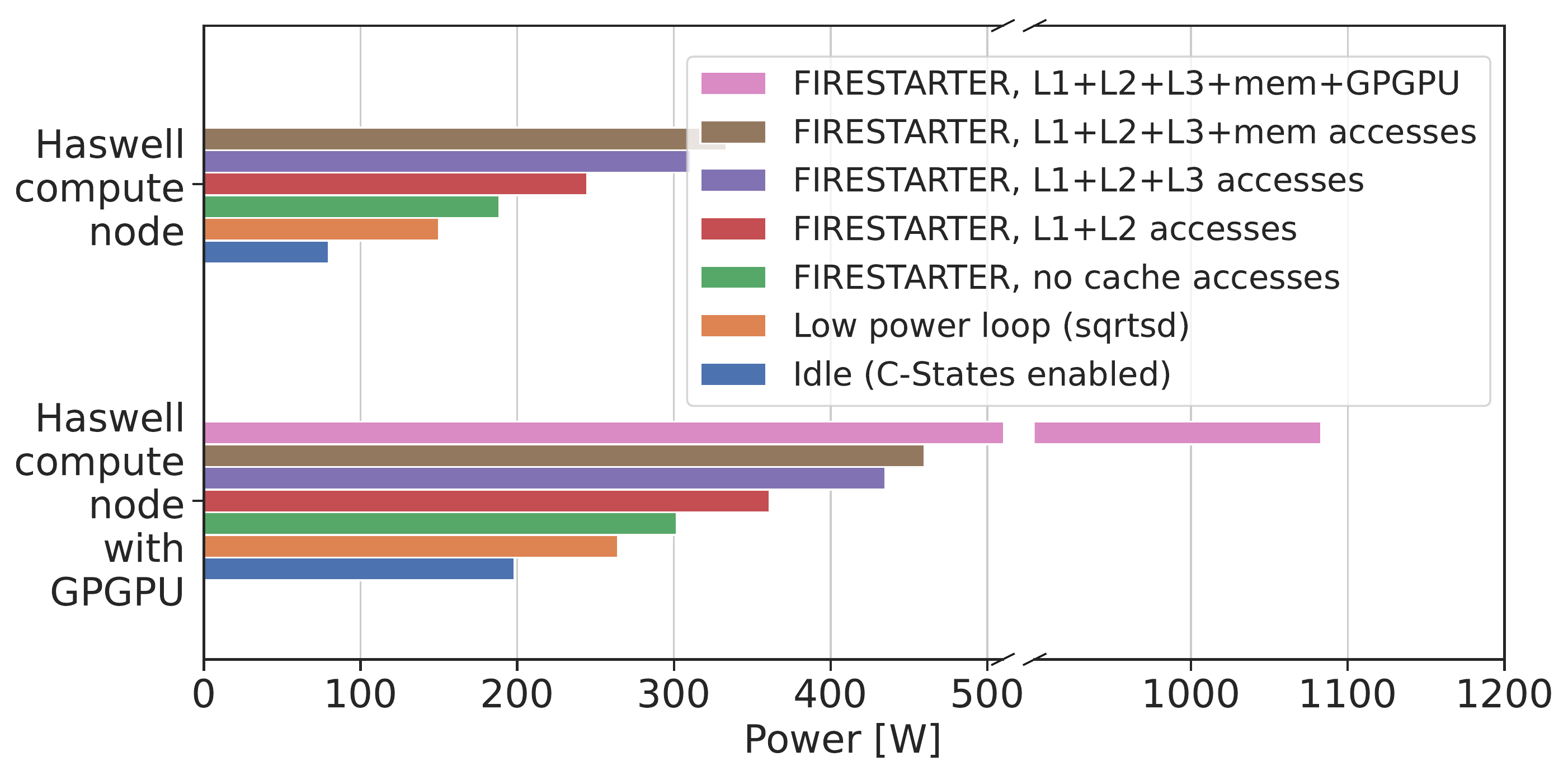}
		\caption{\label{fig:gpu2power}FIRESTARTER~2 optimized for maximum power consumption with different cache accesses on two systems with 2x Intel Xeon E5-2680 v3 each (at \SI{2000}{\MHz} to avoid throttling based on AVX-frequencies), one with 4x NVIDIA K80. Each value represents mean power over \SI{240}{\second}, excluding initial \SI{120}{} and final \SI{2}{\second}. Each memory level adds to total power consumption. Each GPGPU adds \SI{29}{\watt} (idle) to \SI{156}{\watt} (stress).}
	\end{figure}

	Power consumption is one of the limiting factors in processor design, especially since the end of Dennard Scaling~\cite{Dennard_07_Scaling}.
	It consists of two parts: static and dynamic power consumption~\cite[Section 5.1.3]{Weste_2010_CMOS_VLSI}.
	The former defines the power needed to retain the internal state of the processor.
	The latter is the additional power to operate, e.g., to process a specific instruction on an execution unit.
	Over the years, manufacturers implemented different power-saving strategies to lower both.
	This includes Dynamic Voltage and Frequency Scaling (DVFS)~\cite{dvfs}, clock gating~\cite[Section 5.2.1.1]{Weste_2010_CMOS_VLSI}, and power gating~\cite[Section 5.3.2]{Weste_2010_CMOS_VLSI}.
	These require to disable components or lower the processor performance and are typically not used in fully occupied systems where all processor cores are active and executing code.
	In such active scenarios, power consumption highly depends on the workload~\cite{Molka_2010_Energy}, the processed data~\cite{Lucas_2016_AluPower,Schoene_2019_SKL}, and the underlying processor architecture~\cite{Molka_2010_Energy} and its implementation~\cite{Rountree_2021_Performance_Power_Bound}.
	Leading to two different scenarios: either processor power at reference frequency is lower than the defined power limits, and Turbo mechanisms can be applied, or the consumption is higher than the limits and frequencies need to be throttled~\cite{Rountree_2021_Performance_Power_Bound, Hackenberg_2015_Haswell, Schoene_2019_SKL}.
	This budgeting introduces uncertainty in processor performance and power dissipation and challenges administrators and system designers.
	To test whether the electrical and cooling infrastructure fit the installed components, operators need to run worst-case workloads on computing nodes that maximize the dynamic power consumption.
	The workloads stress core components such as execution units and caches, off-core components like additional caches and main memory, and accelerators.
	This is depicted in \figref{gpu2power}.

	\subsection{Processor Stress Tests}
	\label{sec:related}
	Power stress tests usually consist of specific workloads, which vary depending on the targeted components and implementation details.
	In the following, we summarize popular stress tests listed in \tabref{overview}.
	
	\textit{Prime95}~\cite{mprime_website} tests Mersenne numbers for primality using the Lucas-Lehmer-Test.
	It makes use of the available SIMD extensions and uses parallel fused multiply-add instructions.
	While Prime95 reaches high power consumption, it has to be configured carefully and can show varying power consumption over time as Hackenberg et al. show in~\cite{Hackenberg_2013_FIRESTARTER}.
	It also provides an error check, which allows users to also test the computational stability of their system.
	\newpage
	The \textit{LINPACK} benchmark~\cite{linpack} is a numerical library to solve linear equations.
	While there are different implementations to match different architectures, e.g., for shared and distributed memory systems, the main algorithm is the same.
	It makes use of dense matrix algorithms and mainly floating point multiply-add instructions for computations, which exist in various SIMD instructions sets.
	In contrast to other benchmarks, the High Performance Linpack is parallelized via MPI and is therefore also applicable to distributed memory systems.
	Like Prime95, it also checks whether the result of the computation is correct.
	However, reoccurring initialization and finalization phases can significantly lower power consumption.
	
	\textit{stress-ng}~\cite{stressng_presentation,stressng_website} is a Linux stress testing utility that includes various workloads to stress different aspects of a system.
	These workloads range from numerical (e.g., searching, sorting) to hardware (e.g., sqrt, stream) and operating system (e.g., fork, malloc).
	While a variety of tests is available, none of them maximizes power consumption.
	The matrix-matrix product could result in a high power consumption but it currently\footnote{git hash cd92897e7932775009479fedd5da605d4e1ce462} uses long doubles, which are not supported by SIMD extensions.
	The code is also written in C, and the compiler would need to vectorize it automatically.
	For some workloads, stress-ng supports result verification, but that is not the default setting.
	
	\textit{eeMark}~\cite{Molka_12_eeMark} is an artificial stress test targeted at distributed memory systems.
	It consists of compute, network (MPI), and I/O routines, which can be linked together.
	The concrete workload is chosen by selecting templates which define the used datatype, and the type and order of routines, which are compiled into a benchmark.
	The compute-intense workload is defined by a blocked matrix-matrix multiplication for single and double precision data written in C, which requires an optimizing compiler to vectorize the code.
	While eeMark guarantees that each routine will result in valid data (i.e., no denormals or zeros), bit-flips can still alter the results, which are not checked.
	
	The \textit{FIRESTARTER}~\cite{Hackenberg_2013_FIRESTARTER} stress test is based on assembly kernels that target specific processor components.
	These kernels are specialized for specific x86\_64 architectures and stress mainly the SIMD execution units and the memory architecture.
	FIRESTARTER also supports NVIDIA Graphics Processing Units (GPUs).
	Designed as an out-of-the-box solution, it does not need to be manually configured but still results in continuous high power consumption.
	Nevertheless, it has weaknesses, which we further discuss and optimize in the following sections.

	\newpage
	\section{FIRESTARTER -- Design}
	\label{sec:firestarter2}
	To achieve the goal of maximizing power consumption and thermal output, FIRESTARTER consist of multiple workloads $\omega_k \in \Omega$ that stress specific components.
	Each consisting of a loop based on the following variables:
	\begin{itemize}
		\item The used \textbf{set of instructions $I$}, consisting of specific instructions that are executed,
		\item The \textbf{unroll factor $u$}, defining the number of instructions in the workload loop, and
		\item The \textbf{memory accesses $M$}, defining how operands are accessed from which memory level.
	\end{itemize}
	\begin{figure}[t]
		\centering
		\includegraphics[width=\columnwidth]{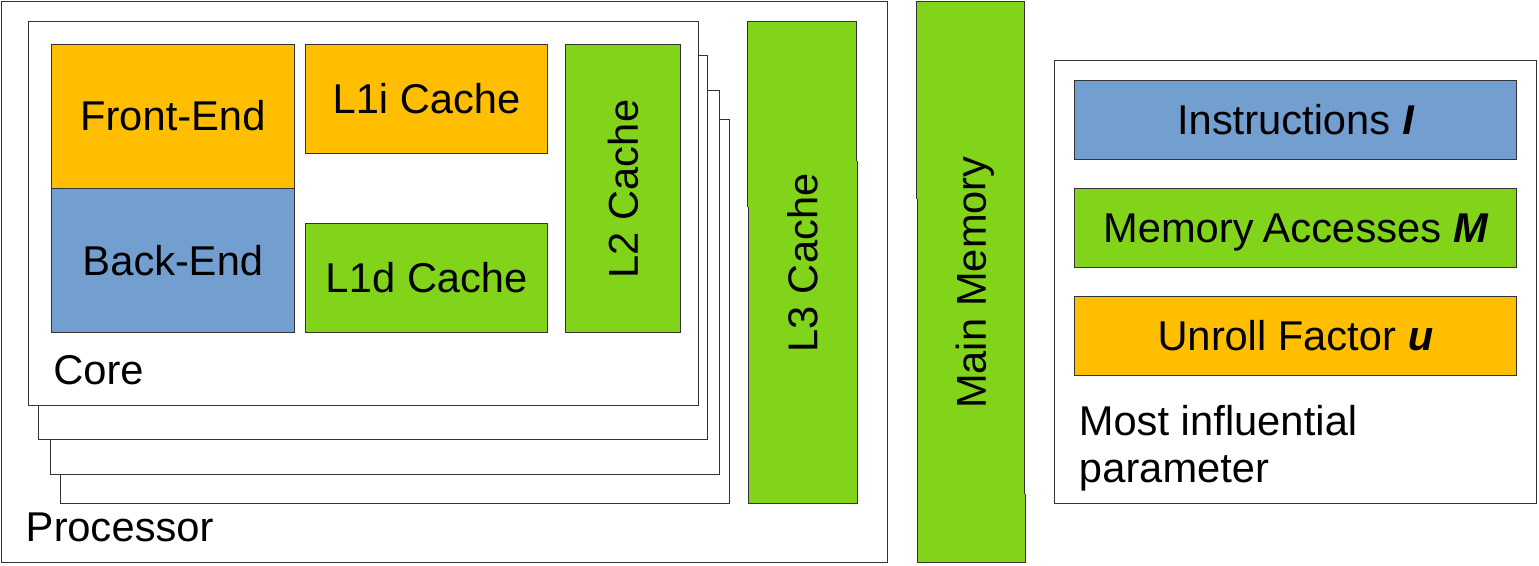}
		\caption{\label{fig:components}Stressed components within a processor and the most influential parameter.}
	\end{figure}
	\figref{components} depicts which parts of a computer are stressed most by which of the factors.
	The target of the instructions within $I$ is to increase power consumption by running complex operations and increase instruction parallelism.
	Therefore, the workload typically uses the widest supported SIMD-Floating-Point-instructions with the highest complexity (Fused Multiply-Add, FMA if available) that can run in a pipelined mode without any stalls.	
	Additionally, $I$ contains integer instructions, which increases parallelism and power consumption further.
	However, each set of instructions can be performed in a single cycle on its target platform.
	A list of these can be retrieved with the \texttt{-a}/\texttt{--avail} parameter.
	A set of instructions can be chosen via the \texttt{-i}/\texttt{--function} parameter.
	FIRESTARTER utilizes $u$ to ensure that the processor front-end can continuously decode the instructions and is not clock-gated.
	Hence, the loop should hold enough instructions to not fit into the typical (\textmu)Op Buffers and Loop-Stream Buffers of processors.
	However, the size of the inner loop should also not exceed the size of the L1-Instruction-cache to avoid adding instruction fetch stalls.
	$u$ is an integer that defines how often the given instructions should be unrolled within the inner loop.

	The last factor is memory accesses $M$.
	It specifies how the memory hierarchy is utilized (\underline{REG}isters, caches, \underline{RAM}), the access pattern for non-register levels (\underline{L}oad, \underline{S}tore,\underline{L}oad+\underline{S}tore, \underline{2 L}oads+\underline{S}tore, \underline{P}refetch)\footnote{Not all patterns are defined for all levels.} and the occurrences within the given loop as an integer $a$ as defined in Equation~\ref{eqn:M}.

\footnotesize
\begin{equation}
 M \subseteq \left\{
  \left(
  \text{REG} \cup
  \left(
    \left\{
      \begin{array}{c}
        \text{L1\_} \\
        \text{L2\_} \\
        \text{L3\_} \\
        \text{RAM\_} \\
      \end{array}
    \right\}
    \times
    \left\{
      \begin{array}{c}
        \text{L} \\
        \text{S} \\
        \text{LS} \\
        \text{2LS} \\
        \text{P} \\
      \end{array}
    \right\}
  \right)
  \right)
  \times
  \left(
    a \in \mathbb{N}^+
  \right)
\right\}
\label{eqn:M}
\end{equation}
\normalsize
	They are defined as a set of access definition and $a$, e.g., \texttt{REG:4,L1\_L:2,L2\_L:1} for accesses to registers and loading access to L1 data and L2 cache.
	The different numbers $a_i$ of the multiple accesses $m_i \in M$ are combined in the following way: Based on the fraction of $a_i$ in the total number of all defined accesses ($\sum a_i$), the unrolled sets of instructions perform the accesses based on the occurrences.
	In the previously given example (\texttt{REG:4,L1\_L:2,L2\_L:1}), from seven consecutive sets of instructions in the inner loop, four will work within registers, two will access L1 and one will access L2.
	Within the consecutive accesses, the single entries will be distributed as good as possible so that the L1 accesses will have a distance of at least three sets of instructions.
	The consecutive accesses are then unrolled so that the total number of instruction sets equals $u$.

	Memory accesses have to be chosen carefully.
	On one hand, under-utilization can be a problem if there are not enough of them.
	On the other, if there are too many accesses and the out-of-order engine cannot cover the memory latencies, processor stalls will occur.
	Both effects can lower power consumption.
	
	\begin{figure}[t]
		\centering
		\includegraphics[width=\columnwidth]{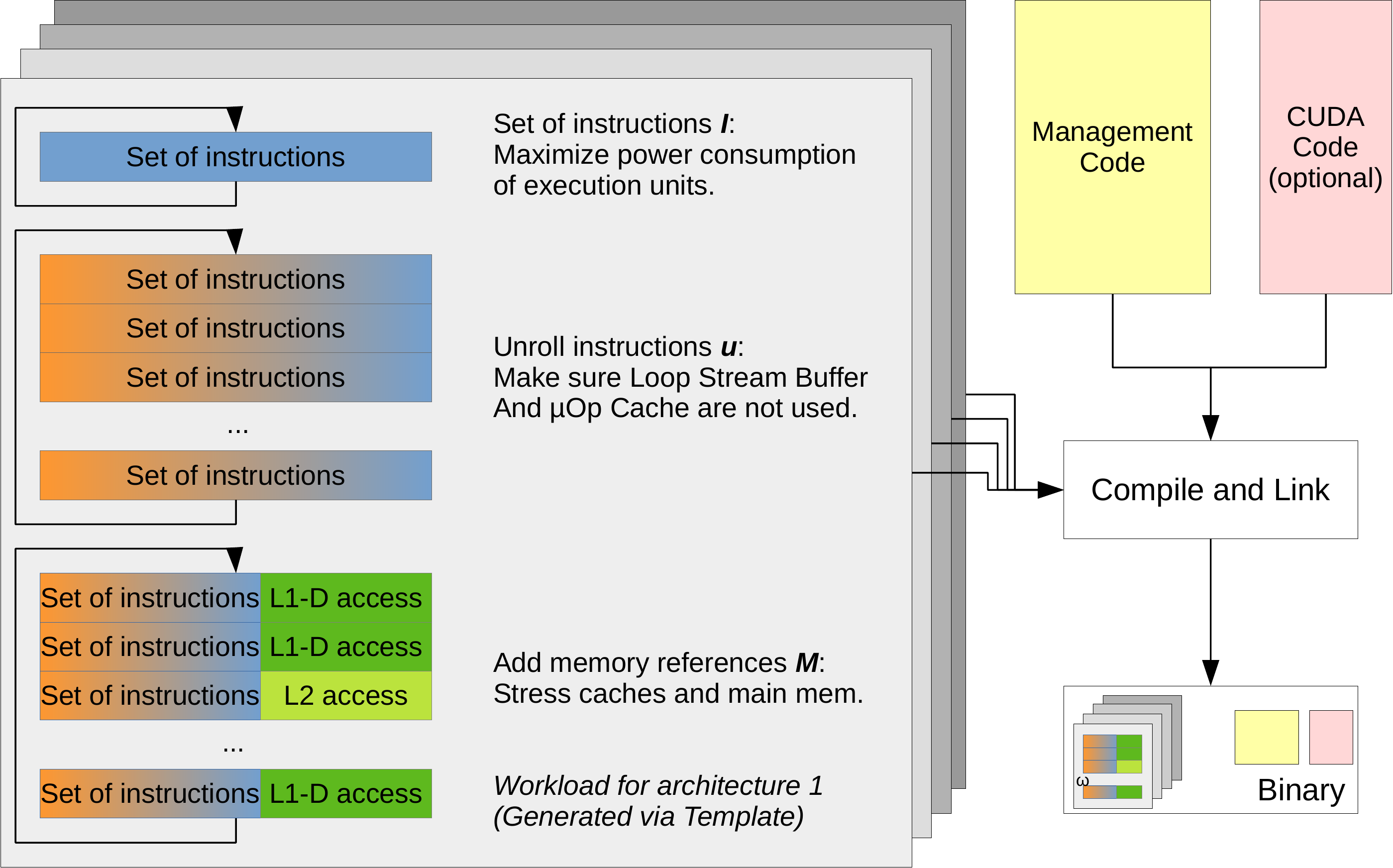}
		\caption{\label{fig:fs1-compile}Compilation process in FIRESTARTER 1.x: Different workloads are generated for different processor models based on instruction mix definitions, unroll factor, and memory accesses. These are compiled and linked to the binary together with management information and CUDA code.}
\vspace{8mm}
	\end{figure}
	\subsection{Challenges}
	Previous versions of FIRESTARTER held a fixed set of available workloads, each optimized for a specific Stock Keeping Unit (SKU) of an x86\_64 architecture.
	These workloads $\{\omega_1, \omega_2, ..., \omega_n\}$ are defined using templates and have a fixed set of parameters $I_k$, $u_k$, and $M_k$ with $1\le k\le n$.
	The build process generates C files from these templates and later compiles them into the binary.
	\figref{fs1-compile} shows the compilation process.
	During execution, FIRESTARTER checks for CPU vendor, family, and model and runs the workload $\omega_k$ for the best fitting processor.

\newpage
	However, this static approach of using an SKU-optimized workload does not necessarily work for other SKUs of the same family and model: a different number of cores and different core frequencies significantly influence how off-core components can be used without introducing stalls.
	Different processors of \textit{the same SKU} can experience a sub-optimal workload, for example, due to DRAM timings.
	Depending on the installed memory modules, memory bandwidth and latency can significantly differ.
	Hence, a different set of memory accesses $M$ is necessary to avoid stalls.
			
	\begin{figure}[t]
		\centering
		\includegraphics[width=.65\columnwidth]{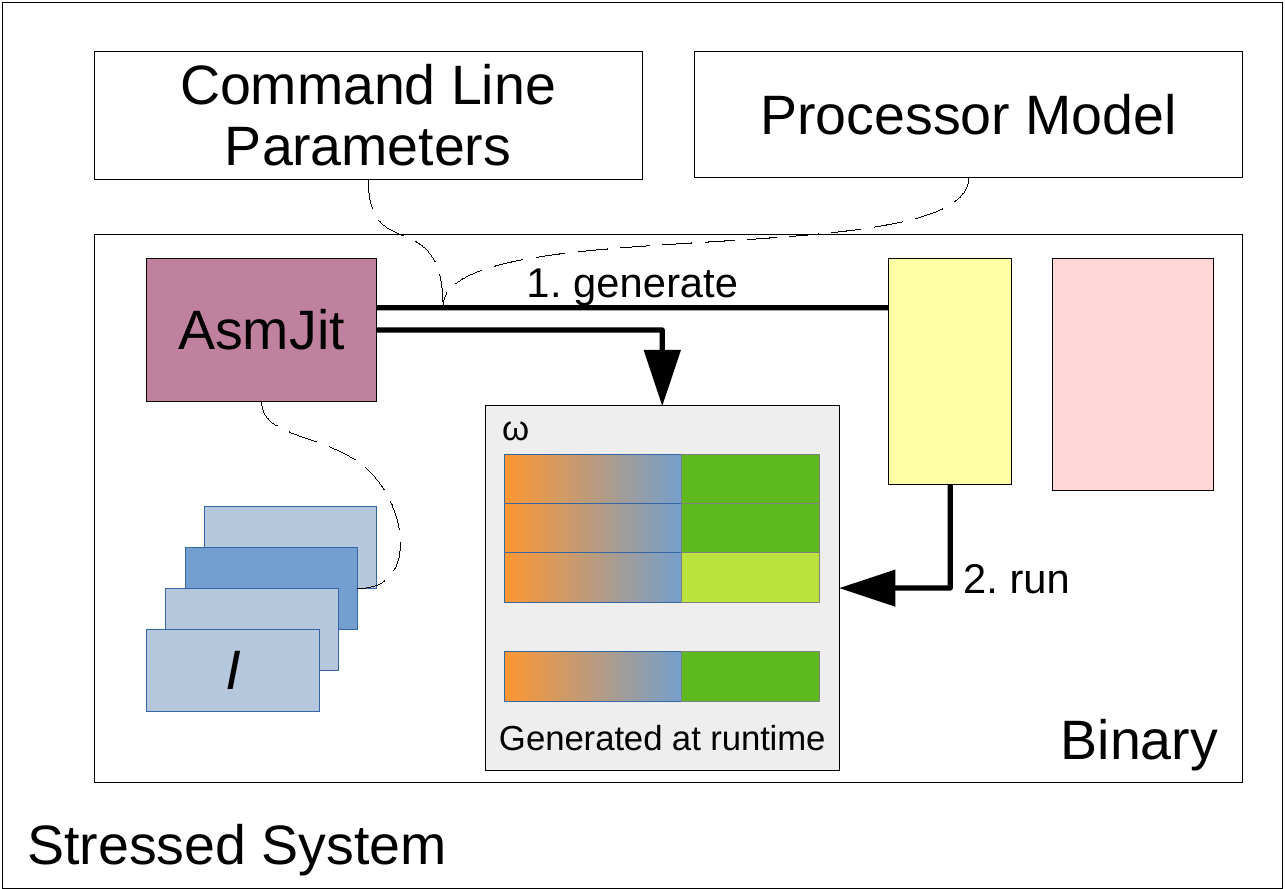}
		\caption{\label{fig:fs2-run}Runtime process in FIRESTARTER~2: The binary checks the processor model of the current system and uses AsmJit to create a matching workload. Memory accesses and unroll factor can also be set via command line parameters or found using automated tuning as described in \secref{auto-tuning-implementation}.}
	\end{figure}
	
	\subsection{Online Workload Creation}
	\label{sec:asmjit}
	In previous versions of FIRESTARTER, finding an optimal configuration for $M$ was a tedious task.
	To alter a workload taking different DRAM timings and a different number of processor cores into account, users of FIRESTARTER had to (1) recreate the source code by changing template parameters, (2) compile that source code, and (3) test the created binary.
	If they were not satisfied with the result, they had to repeat the procedure.
	To find an optimum, which maximizes power consumption, the processor stressing workload would be interrupted regularly for re-compilation.
	The lower power consumption during preparation and compilation also requires a new warm-up phase for measuring the power consumption of the workload.
	
	To speed up and simplify this procedure and to lower the dependencies on external packages for such a task, we introduced the Just-In-Time compiler AsmJit~\cite{Kobalicek_AsmJit} into FIRESTARTER~2.
	Now, the binary carries only the instruction mix definitions but not the concrete representation of the workloads.
	Users can define the unroll factor\footnote{Argument \texttt{---set-line-count}} $u$ and the memory accesses\footnote{Argument \texttt{---run-instruction-groups}} $M$ at runtime.
	FIRESTARTER uses these runtime parameters to create the binary representation of the workload, which is then executed as shown in \figref{fs2-run}.
	To ensure backward compatibility, FIRESTARTER~2 also supports the previous workload definitions.
	The default unroll factor $u$ is set to a default value to ensure instructions are fetched from the L1-Instruction-cache.
	We explicitly exclude $I$ from tuning since user-defined instructions that are chosen poorly increase the chances of producing overflows, underflows, denormalized values, and other unwanted behavior, which could lower power consumption.
	
	\subsection{Automatic Tuning}
	\label{sec:auto-tuning-implementation}
	We included an internal optimization and metric measurement loop that tunes the memory accesses within $M$ to achieve high power consumption.
	This optimization loop takes into account various performance metrics as well as power measurements.
	
\begin{figure}[t]
		\centering
		\includegraphics[width=\columnwidth]{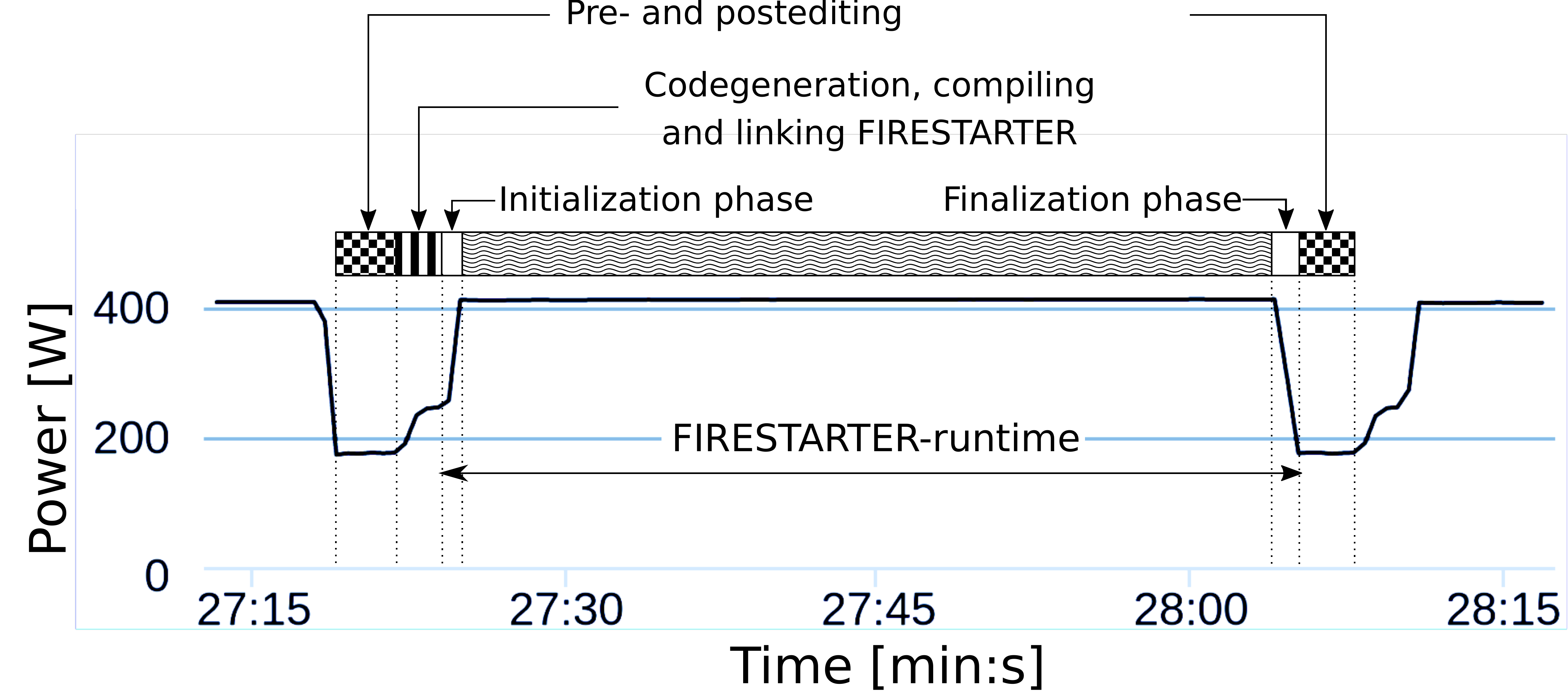}
		\caption{\label{fig:runtime_ignite} Overlay of an actual power measurement and the different phases of a single optimization iteration in the FIRESTARTER 1.x automatic tuning prototype~\cite{Hoehlig_2016}.
		}
	\end{figure}

\begin{figure}
		\centering
		\includegraphics[width=\columnwidth]{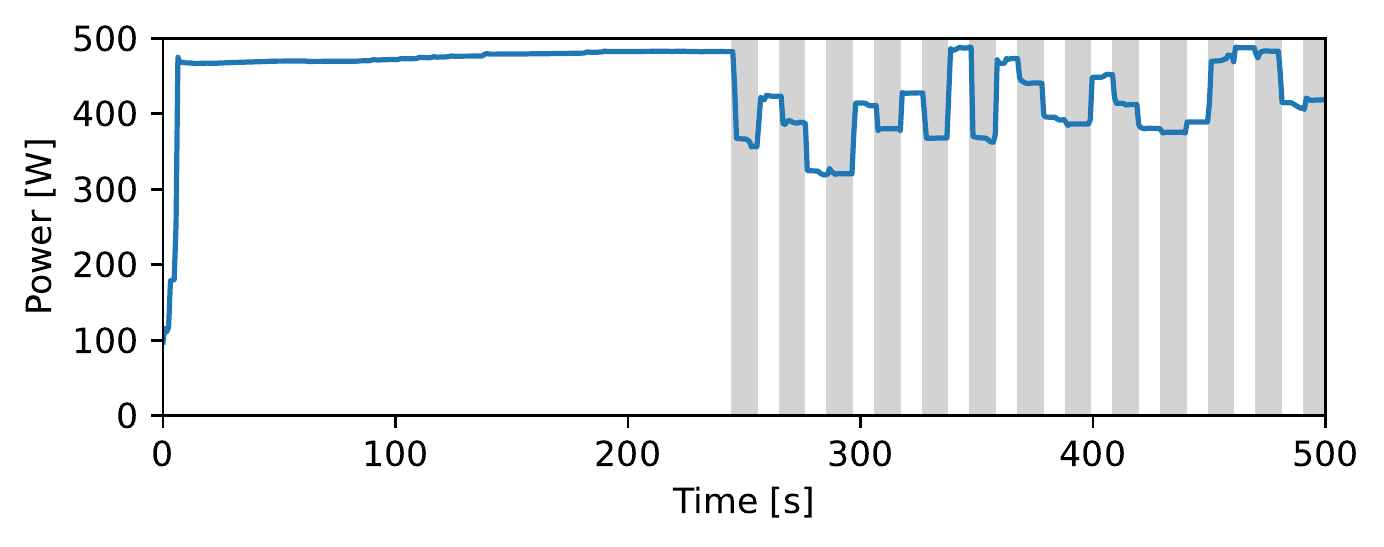}
		\caption{\label{fig:runtime_firestarter2}Overlay of an actual power measurement and the different phases of FIRESTARTER~2s automatic tuning. First, the processor is preheated for \SI{240}{\second}. Then, workload candidates are tested (shaded gray and white areas). Compared to \figref{runtime_ignite} the measurement duration is significantly reduced and there is no visible drop in power consumption between candidates. Only the start of the optimization is depicted.}
	\end{figure}

	In his master thesis, H{\"o}hlig~\cite{Hoehlig_2016} proposed a prototype using evolutionary algorithms.
	However, with the previous FIRESTARTER design, the program needed to be recompiled each time the parameters changed.
	Like the manual optimization, it resulted in the same suboptimal runtime properties -- a few seconds of lower power consumption between each test run and thereby a longer measurement duration of a few minutes rather than seconds to mitigate thermal effects.
	Figure \ref{fig:runtime_ignite} shows this behavior.
	With the new design and the runtime generation of workloads, we can speed up this process as depicted in \figref{runtime_firestarter2}.
	
	\newpage
	While there are multiple possible optimization algorithms, there is an additional constraint.
	One design goal of FIRESTARTER is the ease of use.
	This goal includes removing dependencies and using only statically linked libraries, limiting the options of possible optimization functions.
	Therefore, we opted for NSGA-II~\cite{Deb_2002_NSGA-II}, a well-known and easy-to-implement multi-objective algorithm.
	Compared to other non-dominated multi-objective sorting-based evolutionary algorithms, it does not need a sharing parameter and has a runtime complexity of only $O(MN^2)$, where $M$ is the number of objectives (in our case the number of accesses to the various memory levels), and $N$ is the population size.

	NGSA-II runs a multi-objective optimization and converges near the Pareto-optimal set.
	To do so, FIRESTARTER~2 collects two metrics (objectives) during the execution of the compute kernel, typically power consumption and instruction throughput.
	By default, it supports three built-in metrics:
	First, measuring the average power consumption over time with the Intel Running Average Power Limit (RAPL) mechanism via the \texttt{sysfs} interface provided by the Intel RAPL kernel module.
	This is the most convenient way for users to measure power on Intel systems and is accurate for the measured domains since the Haswell generation of processors~\cite{Hackenberg_2015_Haswell}, while it is less accurate on older Intel~\cite{Hackenberg_2013_Power} and newer AMD systems~\cite{Schoene_2021_Zen2}.
	Second, measuring instructions per cycle (IPC) using the \texttt{perf\_event\_open} syscall\footnote{\url{https://man7.org/linux/man-pages/man2/perf_event_open.2.html}}.
	We chose the IPC metric to ensure that the optimization process is not stuck in local maxima, where only parts of the system are used.
	Furthermore, this metric supports finding a per-architecture optimum: Ignoring the instruction throughput could favor workloads that use too many memory accesses for SKUs with higher frequencies and core counts.
	However, the used hardware performance events (executed instructions, processor cycles) are available on most processors and can monitor whether additional memory accesses lower the instruction throughput and the applied frequency.
	Finally, we also integrate an IPC estimation metric, which is valuable if the syscall is not available or access rights restrict the availability of its functionality.
	This estimation is based on an assumption of a constant frequency and a counting of the executed inner loops.
	However, this approach is distorted if the frequency of the processor changes during the optimization run.

	In addition to the integrated metrics, users can create custom metrics via external binaries, scripts, and libraries.
	A simple Python script could, for example, forward power measurement values from an external power meter.
	Alternatively, libraries written in C/C++ can provide the same functionality with less overhead.
	The integration of the tuning cycle leads to a faster convergence to high power consumption and the elimination of near-idle phases between single iterations of the optimization process.
	\subsection{Other Changes}
	In \cite{Schoene_2019_SKL}, Schöne et al. demonstrated that the data of computations have a significant impact on power consumption.
	Lucas et al. showed this effect on the arithmetic logic unit on GPUs~\cite{Lucas_2016_AluPower}.
	Hickmann et al.~\cite{hickmann_bradford_fletcher_2016} filed a patent assigned to Intel describing means to clock-gate specific parts of the FMA-Unit when \mycite{an answer is either trivially known or can be computed in a more power-efficient way}.
In particular, the operands should not equal $+\infty$, $-\infty$, or $0$, as these would make the computations trivial, thereby dramatically reducing the power consumption.
This knowledge adds another design restriction for instruction mixes.
Due to a bug in previous versions, values in registers accumulated to $+\infty$ and $-\infty$, resulting in a lower power consumption.
To demonstrate this behavior we run version 1.7.4 and 2.0 without any memory references\footnote{with appropriately modified config file before compiling or the command-line argument \texttt{---run-instruction-groups=REG:1}, respectively}.
The measurement was run on our test system described in \tabref{test-system} at nominal frequency.
It took the average over a runtime of 240 seconds, excluding the initial 120 and last 2 seconds.
The new version has a higher power consumption with \SI{314.1}{\watt} compared to the older version with \SI{305.6}{\watt}.
\begin{table}[t]
	\centering
	\caption{\label{tab:test-system}Test system details}
	\begin{tabular}{rr}
		\toprule
		Processor	&	2x AMD EPYC 7502 \\
		\rowcolor[HTML]{EFEFEF}Cores		&	2x 32 \\
		Available frequencies	&	\SI{1500}, \SI{2200}, \textbf{\SI{2500}{\mega\hertz}} (nominal), Turbo \\
		\rowcolor[HTML]{EFEFEF}L1-I and L1-D cache	&	64x \SI{32}{\kibi\byte} + \SI{32}{\kibi\byte} \\
		L2 cache	&	64x \SI{512}{\kibi\byte} \\
		\rowcolor[HTML]{EFEFEF}L3 cache	&	16x \SI{16}{\mebi\byte} \\
		Mainboard	&	GIGABYTE R282-Z90-00 \\
		\rowcolor[HTML]{EFEFEF}Memory		&	16x SK Hynix HMA82GR7CJR8N-XN  \\
		\rowcolor[HTML]{EFEFEF}&	\SI{1600}{\mega\hertz} \\
		Disk		&	Intel SSDSC2KG960G8 \\
		\rowcolor[HTML]{EFEFEF}OS           &   Ubuntu 18.04 LTS \\
		Kernel          &   5.4.0-47-generic \\
		\bottomrule
		\vspace{6mm}
	\end{tabular}
\end{table}

We also added the possibility to flush register contents in regular intervals to a file.
This enables users to check whether their SIMD units still work correctly when processors are used out of their regular specifications (e.g., in overclocked environments).
Furthermore, developers can check whether source code changes lead to diverging numbers.
Optimization metrics can also be used for measurements, where a list of comma-separated values (CSV) are printed after the execution of the workload.
Here, the values are averaged over the whole runtime, excluding an arbitrary time during the start and end of the measurement run, with a default of \SI{5}{\second} and \SI{2}{\second}, respectively.

To stress NVIDIA GPUs, FIRESTARTER uses the DGEMM routines of NVIDIAs cuBLAS library.
However, the initialization of these matrices was inefficient as they were initialized at the host and then transferred to the GPU.
In the new version, data is initialized directly on the GPU.
 \newpage 
\section{Use-Case AMD Zen 2}
\label{sec:zen2}
\subsection{Architecture}
The AMD Zen 2 architecture uses a modular approach where up to eight Core Complex Dies (CCDs) are attached to an I/O-Die.
The CCDs encapsulate up to two Core Complexes (CCXs)\footnote{On the used tested system, each CCD only holds one CCX.}, which include four processor cores each.
While all cores of a CCX share the \SI{16}{\mebi\byte} L3 cache, each of the cores holds its own \SI{512}{\kibi\byte} L2, and 2x\SI{32}{\kibi\byte} L1 Data and Instruction-cache.
The processor cores now support a native execution of \SI{256}{\bit} wide AVX instructions in four floating-point pipes (2x fma/mul\footnote{The execution of FMA instructions can require support of additional pipes.}, 2x add), three address generation unit pipes and four arithmetic logical unit pipes.
The I/O-die connects up to two of the CCDs with a Memory Controller, and, via a network, the CCDs with each other and the I/O interfaces.
To adjust processor performance and power consumption to the current workload, the I/O die and each core have their voltage and frequency, defined by its performance state (P-state).
The highest frequency of the cores of the CCX defines the frequency of the shared L3-cache.
We present more energy efficiency details of this architecture in~\cite{Schoene_2021_Zen2}.
	
\subsection{Test System and Workload Description}
For our test we use a two-socket system equipped with two AMD EPYC 7502 processors (see \tabref{test-system}).
If not stated otherwise, we enabled both hardware threads.
We use a ZES LMG95 power meter to measure power consumption at \SI{20}{Sa \per \second}.
These values are forwarded to MetricQ~\cite{Ilsche_2019_MetricQ} and then processed within FIRESTARTER.
	
The inclusion of the self-tuning algorithm enables users to create a specific workload targeting their particular system.
In our case study, we use the set of instructions for the Intel Haswell platform~\cite{Hackenberg_2015_Haswell} introduced with FIRESTARTER 1.1.
It consists of a mix of two FMA (\texttt{vfmadd231pd}) and two ALU (\texttt{xor}+shift\footnote{alternating shift left/right (\texttt{shl}/\texttt{shr}) to toggle between states 0b0101$\ldots$01 and 0b1010$\ldots$10}) instructions\footnote{Optional stores replace some instructions with \texttt{vmovapd}s.}, resulting in a maximum of four instructions per cycle to be executed.
This workload fits the four-way decoder, which is the bottleneck that would limit the execution in our case.
Alternatively, we could increase the throughput with a workload that fits into the Op Cache for decoded instructions.

\subsection{Influence of the Front-End and the Unroll Factor $u$}
\begin{figure}[t]
	\centering
	\includegraphics[width=\columnwidth]{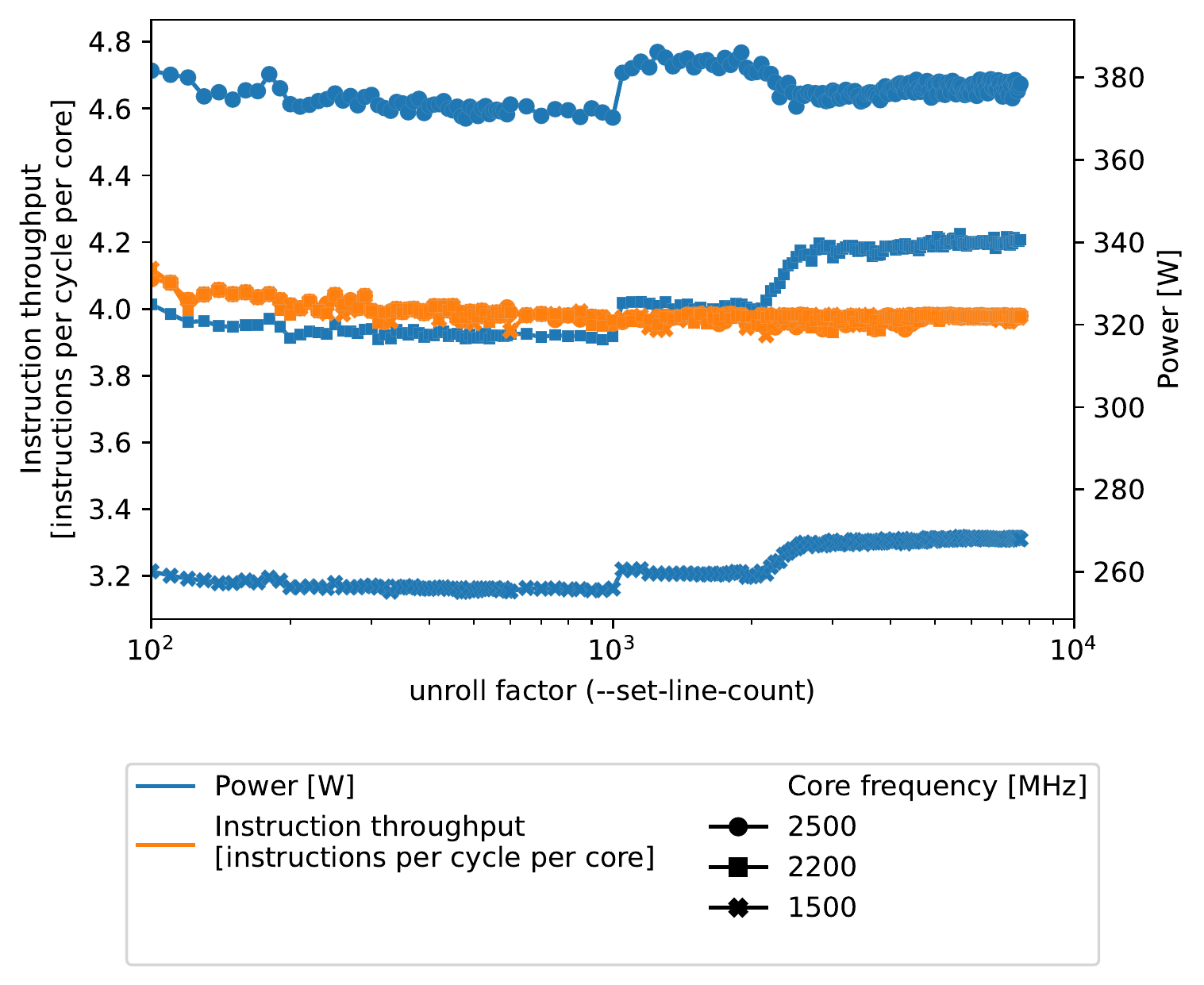}
	\caption{Power consumption and instruction throughput for different unroll factors and P-States.}
	\label{fig:conway_unroll}
\end{figure}

In the first test, we check how the unroll factor influences power consumption.
As described in \secref{firestarter2}, we want to increase power consumption by fetching and decoding instructions.
To test this, we use different unroll factors, which result in three different categories, small, standard, and large, where instructions are loaded from the Op Cache, the L1-I cache, and the L2 cache, respectively.
	We tested for different core frequencies and use a workload that uses loads from the L1-D cache\footnote{\texttt{---run-instruction-groups=L1\_L:1}} to mimic a behavior where memory references are present but not a limiting factor.
	\figref{conway_unroll} shows instruction throughput and power consumption for the tests.
	As expected, the streaming access and a \SI{32}{\byte \per cycle} throughput do not pose a bottleneck.
	As soon as the loop does not fit into the Op Cache any longer (unroll factor 1000), power consumption increases.
	We validated the usage of the Op Cache with the hardware monitoring event 0xAA \textit{UOps Dispatched From Decoder}~\cite[Section 2.1.15.4.4]{AMD_PPR_Server}.
	The instruction throughput does not decrease when instructions have to be served from the L2 cache (unroll factor about 2000).
	However, power consumption increases with the number of caches accessed.
	Surprisingly, it decreases when cores run at a nominal frequency and switch from a normal to a large case, i.e., when accessing L2.
	This can be explained with a frequency throttling that occurs, where the core frequencies are decreased from \SI{2.5}{\giga\hertz} to \SI{2.4}{\giga\hertz} (validated via the hardware monitoring event 0x76 \textit{Cycles not in Halt}~\cite[Section 2.1.15.4.2]{AMD_PPR_Server} referenced by \texttt{perf}s cycle counter)).
	However, in later tests and the final workloads, the L2 cache will be used for loading data instead of instructions.
	To avoid stalls when the L2 cache is used for delivering data and instructions at the same time, we choose the unroll factor so that the loop fits into the L1-I cache.

	\subsection{Influence of the Memory Accesses $M$}
\begin{figure}
		\centering
		\includegraphics[width=\columnwidth]{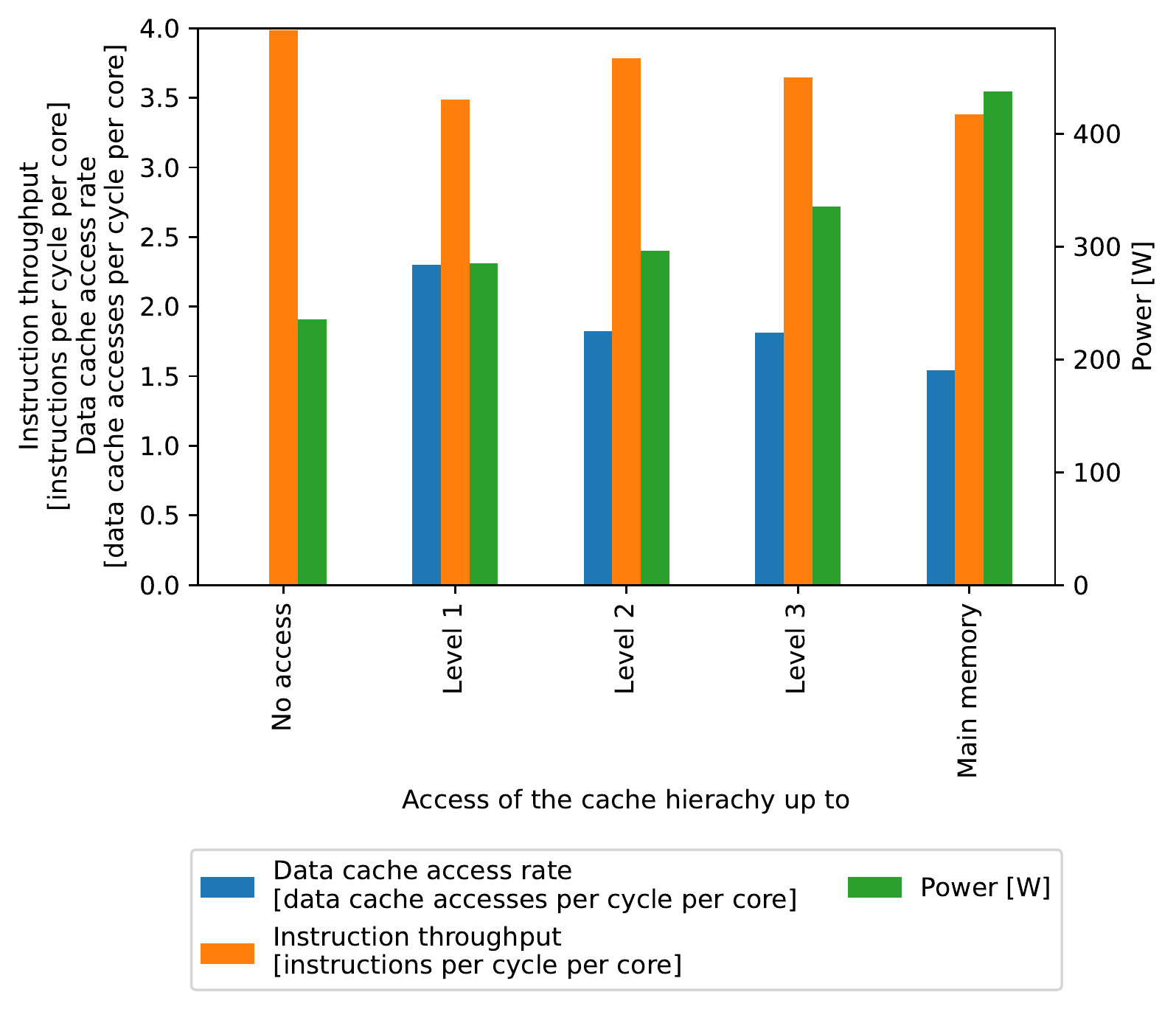}
	\vspace{1mm}
		\caption{\label{fig:conway_memory_accesses}Power consumption, instruction throughput and data cache access rate of FIRESTARTER optimized for accesses to different levels of the cache hierachy on the test system described in \tabref{test-system} (at \SI{1500}{\MHz} to avoid throttling as shown in \secref{tuning-results}). Each measurement took the average over a span of 240 seconds, excluding initial 120 and last 2 seconds.}
	\vspace{3mm}
	\end{figure}

	In a second test, we show how adding memory references $M$ increases the power consumption of the workload.
	To do that, we add accesses to the different cache levels and main memory based on the following strategy:
	We start with the basic FMA-loop, which fits into the L1-I cache using only register operands.
	We then add memory accesses with loads and stores to the L1-D cache for the Level 1 measurement.
	For the remaining measurements, we add loads from additional memory levels.
	To get the ratio with the highest power consumption, we vary the ratio of register calculations and memory accesses.
	We run our test at \SI{1500}{MHz} so that frequency throttling, which was already present in our previous test, is not applied.
	\figref{conway_memory_accesses} shows the results.
	Surprisingly, the number of instructions executed per cycle is reduced in cases where the power consumption is highest.
	However, it only drops to \SI{3.4}{instructions \per cycle}.
	As expected, the power consumption increases with every additionally accessed memory level, leveraging the total consumption from \SI{235}{\watt} to \SI{437}{\watt}, an increase of \SI{86}{\percent}.
	
	\vspace{4mm}
	\subsection{Self-Tuning Results at Different Processor Frequencies}
	\vspace{2mm}
	\label{sec:tuning-results}
	
\begin{figure}[t]
		\centering
		\includegraphics[width=\columnwidth]{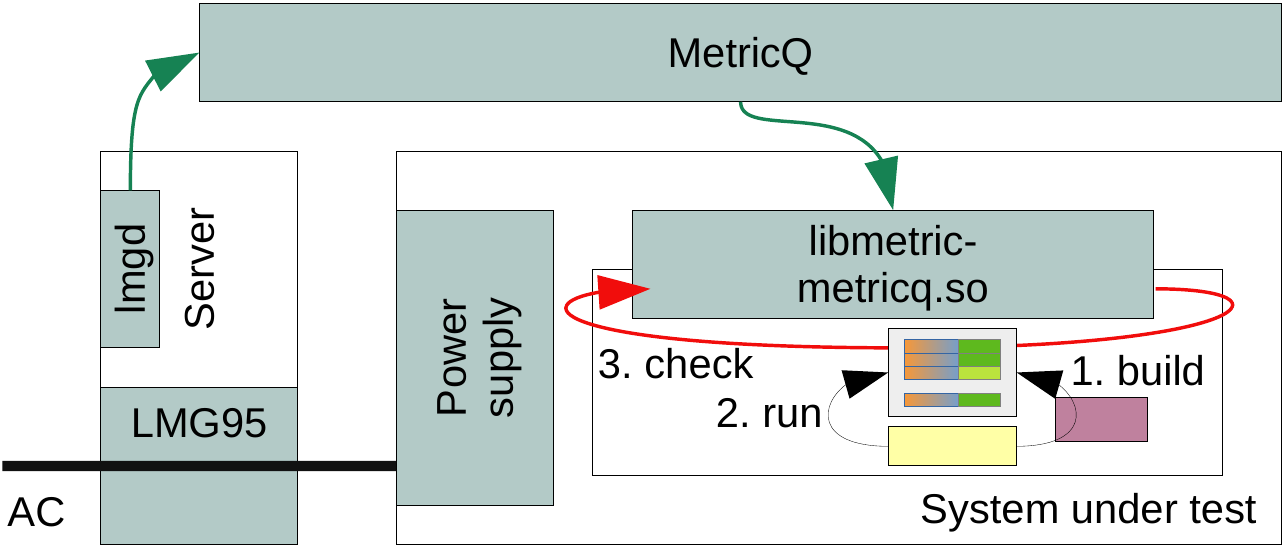}
		\caption{\label{fig:metricq_optimization}Set-up for FIRESTARTER self-tuning: The system power consumption is measured on a remote system. Values are passed to MetricQ where they are buffered. After a workload candidate finished execution, the values are retrieved and processed by the FIRESTARTER and the attached library.}
	\end{figure}

	The final test uses the self-tuning mechanism from \secref{auto-tuning-implementation}.
	\figref{metricq_optimization} shows the overall measurement and tuning infrastructure, and how we integrate power measurement.
	To simulate different processors, we run the optimization at various core frequencies and use the following parameters:

	\begin{itemize}[]
		\item \texttt{---optimize=NSGA2}\\
		defines the optimization algorithm described in \secref{auto-tuning-implementation}.
		\item \texttt{---individuals=40 ---generations=20 ---nsga2-m=0.35}\\ fine-tunes the optimization algorithm.
		In the first generation, an initial population of 40 is randomly initialized and evaluated.
		The following 20 generations are created by binary tournament select, recombination, and mutation (\SI{35}{\percent} probability) from the individuals of the previous generation.
		\item \texttt{-t 10}\\
		specifies the duration for each test
		\item \texttt{---preheat=240}\\
		defines that before the optimization starts, the system runs a default workload for \SI{4}{\minute} to cancel thermal related effects
		\item \texttt{---optimization-metric=metricq,perf-ipc}\\
		names the used optimization metrics
		\item \texttt{---metric-path=libmetric-metricq.so}\\
		specifies the location of the library that is used to retrieve power measurement values
	\end{itemize}
\vspace{8pt}
	When the last generation of workload candidates is evaluated, the best individuals are printed to screen.
	A logfile is saved for further evaluation.
	\begin{figure}[t]
		\centering
		\includegraphics[width=\columnwidth]{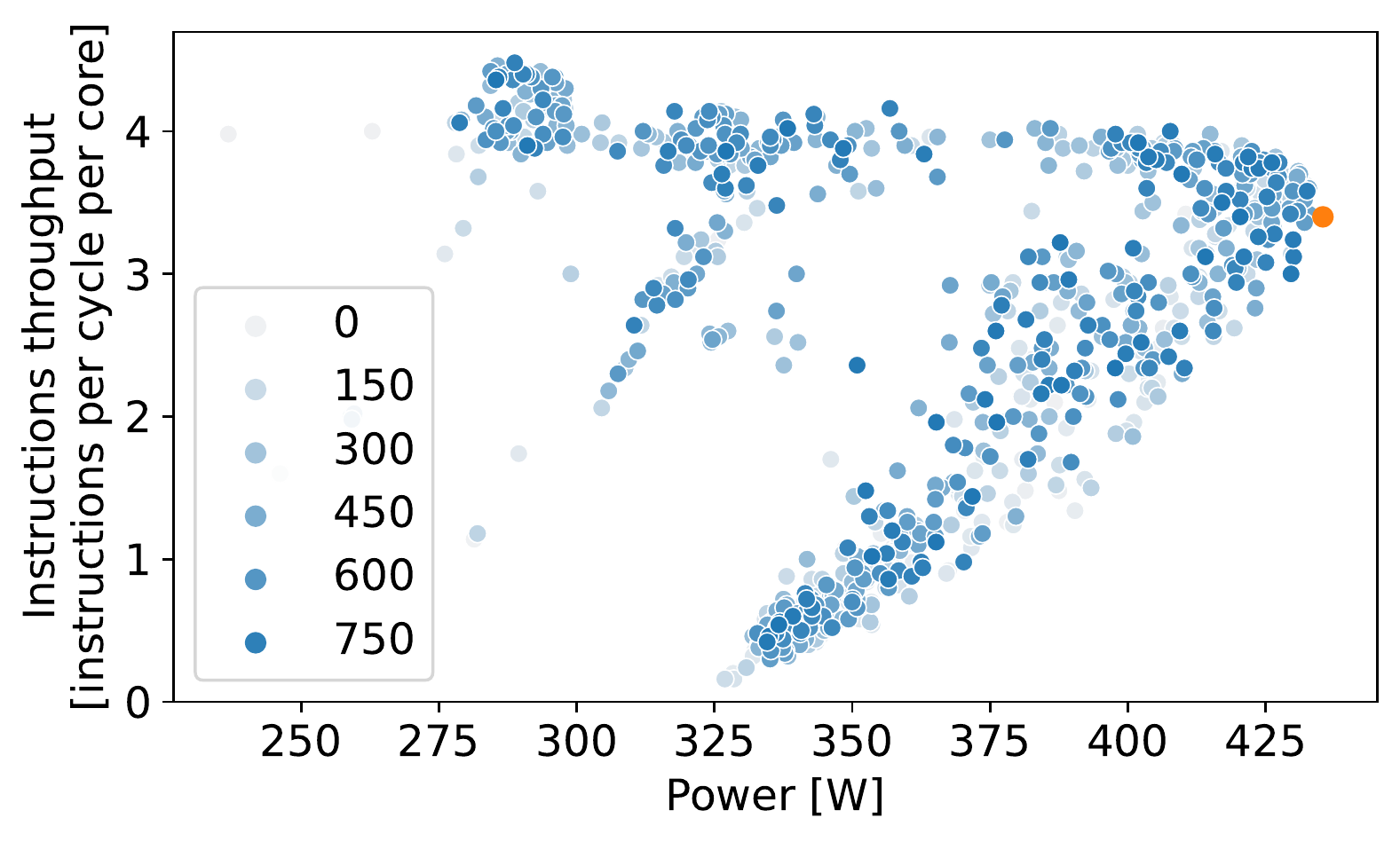}
		\caption{\label{fig:pareto-front}Power and instruction throughput for all evaluated individuals of an optimization at \SI{1500}{\MHz}. The darker the color, the later individuals were evaluated. The selected optimum $\omega_{opt-1500MHz}$ is marked orange.}
	\vspace{2mm}
	\end{figure}
	
	\figref{pareto-front} shows the effectiveness of the optimization by plotting all individuals by both optimization metrics.
	Even though the Pareto front is visible, the optimization still evaluates many individuals in later generations (darker color) inside the hypervolume.
	The final solution meets the target of creating a very high power consumption.
	
\begin{figure*}[t]
\centering
\subfloat[\label{fig:conway-pstate-comparison-power}Power [\SI{}{\W}\symbol{93}]{
	\includegraphics[height=0.195\textheight]{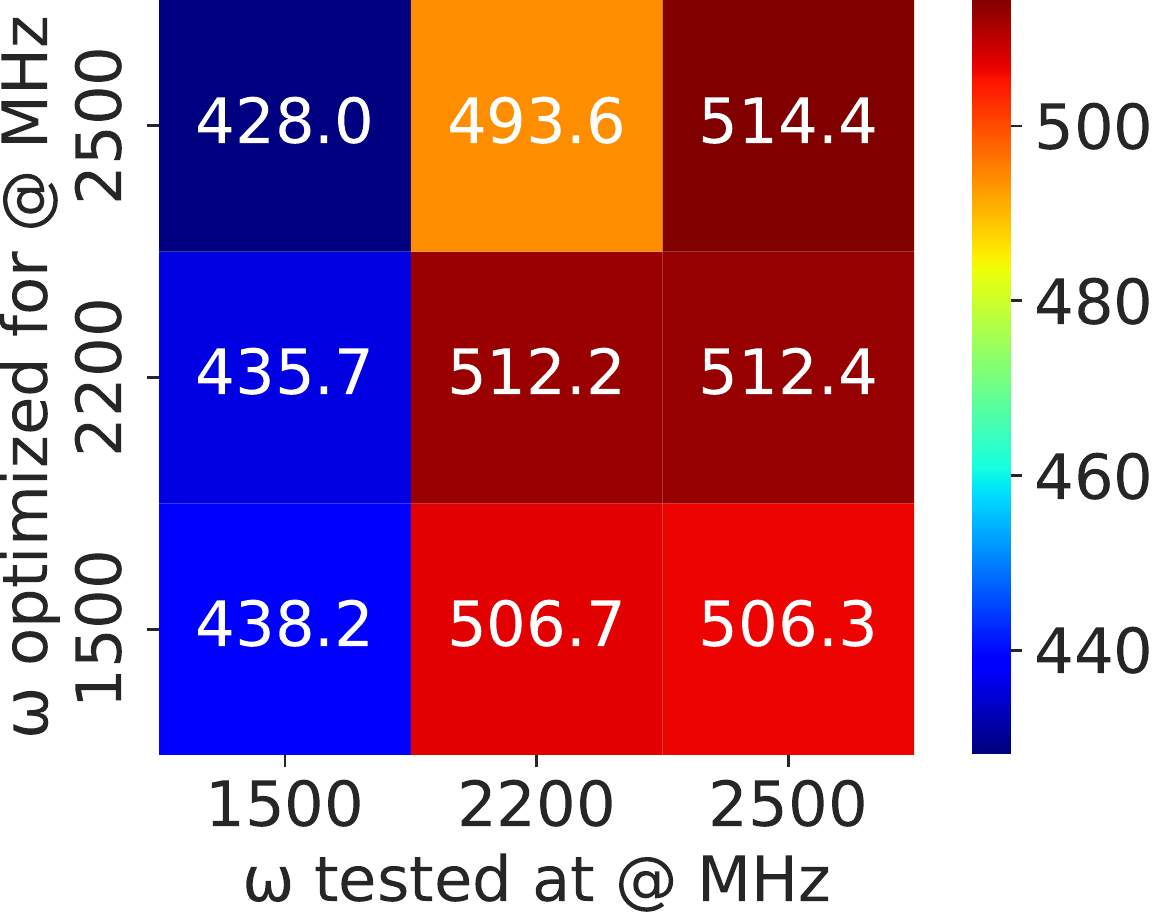}
}
\hfill
\subfloat[\label{fig:conway-pstate-comparison-ipc}Instruction throughput [instructions per cycle per core\symbol{93}]{
	\includegraphics[height=0.2\textheight]{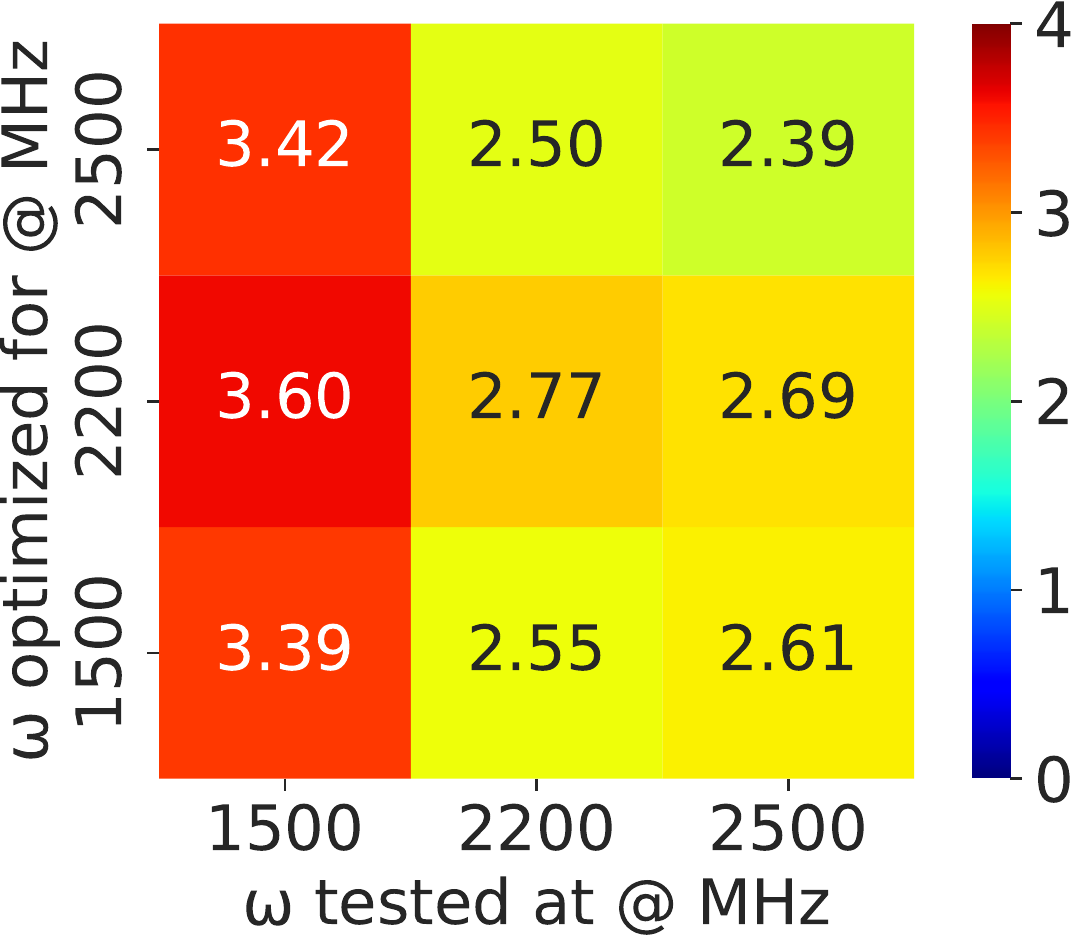}
}
\hfill
\subfloat[\label{fig:conway-pstate-comparison-freq}Core frequency [\SI{}{\MHz}\symbol{93}]{
	\includegraphics[height=0.2\textheight]{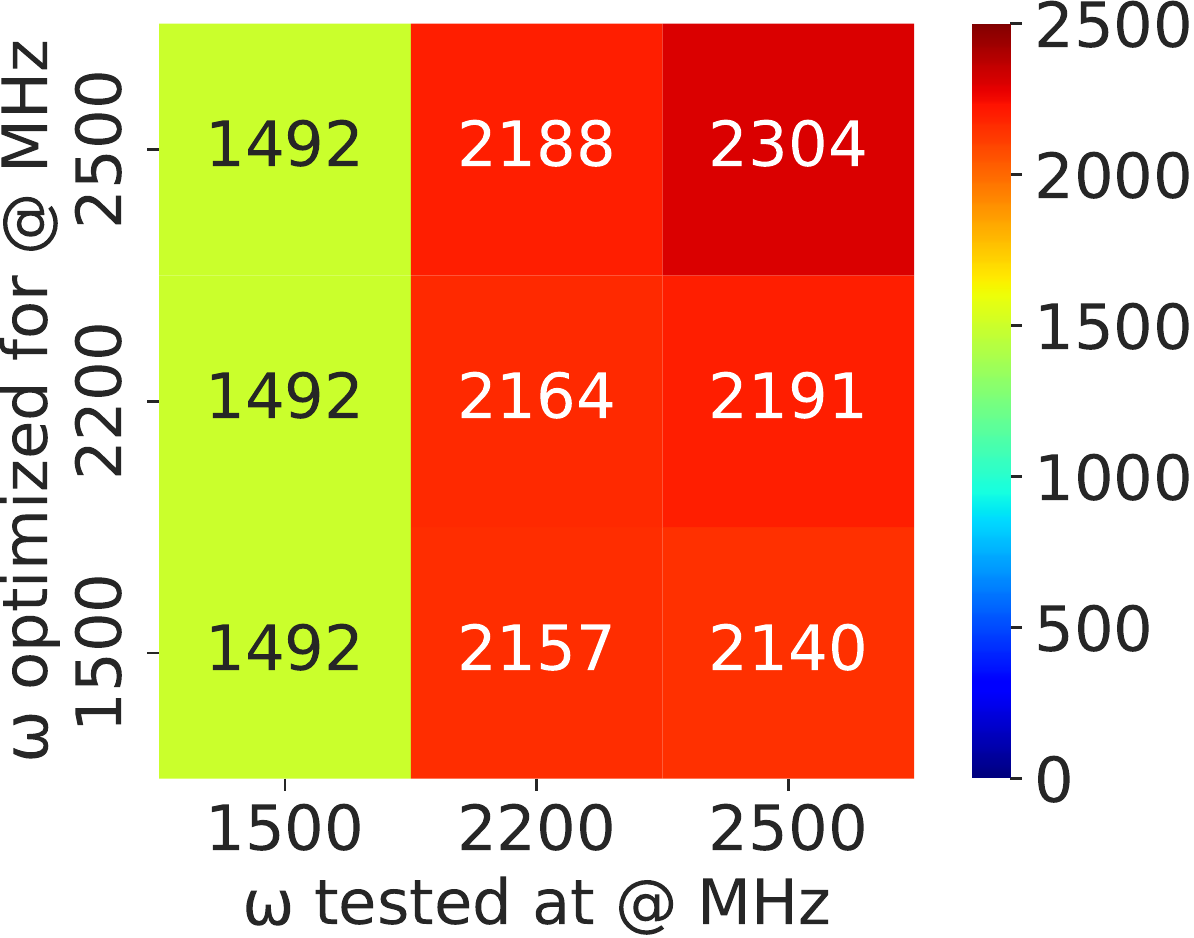}
}
\caption{\label{fig:conway-pstate-comparison}Results for the metrics power, instructions per cycle and frequency for all three optimized versions measured at all three core P-states.
	Each measurement took the average over a span of 240 seconds, disregarding the first 120 and last 2 seconds.
}

	\vspace{10mm}
\end{figure*}
	After generating the three optimized workloads ${\omega_{opt-1500MHz},\omega_{opt-2200MHz},\omega_{opt-2500MHz}}$ for the three different core frequency settings, we evaluate each of them on the available three frequencies.
	We run each test for \SI{240}{\second}, and ignore the first \SI{120}{\second} and final \SI{2}{\second} for the measurement\footnote{\texttt{---measurement -t 240 ---start-delta=120000 ---stop-delta=2000}}.
	During the remaining 118 seconds, we use the internal measurement capabilities of FIRESTARTER and measure the average power consumption of the system, the average instruction throughput over all cores, and the average frequency over all cores.
	\figref{conway-pstate-comparison} shows results for the nine combinations.
	The single heatmaps visualize the performance depending on the core frequency set during the training (y-axis) and the core frequency during the test of the workload (x-axis).
	\figref{conway-pstate-comparison-power} shows that each workload will lead to the highest power consumption for its optimization point (the upward diagonal entries always hold the highest value within the column).
	However, this cannot be said for the instruction throughput, which is affected by the stalls introduced by memory accesses but also by the applied core frequency.
	For example, the workload optimized for \SI{2500}{\mega\hertz} will run at lower frequencies with a higher instruction throughput since the number of stall cycles due to main memory accesses will be reduced.
	However, the workload optimized for \SI{1500}{\mega\hertz} will have a lower instruction throughput when run at \SI{2200}{\mega\hertz} compared to \SI{2500}{\mega\hertz}, since the applied frequency is higher in this case.
	Finally, \figref{conway-pstate-comparison-freq} shows that all three workloads will result in frequency throttling when run at nominal frequency or \SI{2200}{\mega\hertz}.
	Here, the processor decreases its frequency dynamically to avoid peaks that \mycite{cause electrical design current (EDC) specifications to be exceeded}~\cite[Section Floating-Point/Vector Execute]{AMD_Zen2_Overview}.
	
	\section{Conclusion and Future Work}
	\label{sec:summary}
	\vspace{1mm}
	This paper presents a new version of FIRESTARTER, which introduces an optimization loop for creating processor-specific stress-tests.
	We described how FIRESTARTER targets various parts of the system architecture and showed how they contribute to the overall power consumption.
	The newly included self-tuning optimization loop maximizes power consumption and instruction throughput using a multi-objective algorithm.
	We demonstrated the effectiveness of the algorithm on an AMD-ROME-based system.
	We showed that there is no single workflow for a single processor that runs optimal on all possible frequencies.
	All previous versions of FIRESTARTER and the new version are available on GitHub\footnote{\url{https://github.com/tud-zih-energy/FIRESTARTER}} under the terms of the GNU General Public License version 3.
	Future developments will be available in this repository as well.

	For further work, we will include other architectures as provided by ASMJit\footnote{The \textit{initial AArch64 support} has been pushed in 2021-05-18}.
	We also look forward to include the Advanced Matrix Extension (AMX) instructions, which will debut with Intel Sapphire Rapids.
	Finally, we plan to look at support for AMD GPGPUs and specialized GPGPU routines.

	\vspace{3mm}
	\section*{Acknowledgments and Reproducibility}
	\vspace{1mm}
	This work is supported in part by the German Research Foundation (DFG) within the CRC 912-HAEC.
	We want to thank the original developers of FIRESTARTER and the proof-of-concept for the tuning loop: Daniel Molka, Roland Oldenburg, and Stefan H{\"o}hlig.
	Measurement programs, raw data, and chart notebooks are available at \url{https://github.com/tud-zih-energy/2021-firestarter-2}.

	\bibliographystyle{IEEEtran}
	
	\bibliography{paper}

\begin{thebibliography}{10}
\providecommand{\url}[1]{#1}
\csname url@samestyle\endcsname
\providecommand{\newblock}{\relax}
\providecommand{\bibinfo}[2]{#2}
\providecommand{\BIBentrySTDinterwordspacing}{\spaceskip=0pt\relax}
\providecommand{\BIBentryALTinterwordstretchfactor}{4}
\providecommand{\BIBentryALTinterwordspacing}{\spaceskip=\fontdimen2\font plus
\BIBentryALTinterwordstretchfactor\fontdimen3\font minus
  \fontdimen4\font\relax}
\providecommand{\BIBforeignlanguage}[2]{{%
\expandafter\ifx\csname l@#1\endcsname\relax
\typeout{** WARNING: IEEEtran.bst: No hyphenation pattern has been}%
\typeout{** loaded for the language `#1'. Using the pattern for}%
\typeout{** the default language instead.}%
\else
\language=\csname l@#1\endcsname
\fi
#2}}
\providecommand{\BIBdecl}{\relax}
\BIBdecl

\bibitem{bit-flips}
L.~B. Kish, ``{End of Moore's law: thermal (noise) death of integration in
  micro and nano electronics},'' \emph{Physics Letters A}, 2002,
  \href{https://doi.org/10.1016/S0375-9601(02)01365-8}{DOI:
  10.1016/S0375-9601(02)01365-8}.

\bibitem{acpi}
\BIBentryALTinterwordspacing
Y.~Rekhter and T.~Li, ``{ACPI Specification Version 6.4},'' {Unified Extensible
  Firmware Interface Forum}, Tech. Rep., Jan 2021. [Online]. Available:
  \url{https://uefi.org/sites/default/files/resources/ACPI_Spec_6_4_Jan22.pdf}
\BIBentrySTDinterwordspacing

\bibitem{Hackenberg_2013_FIRESTARTER}
D.~{Hackenberg}, R.~{Oldenburg}, D.~{Molka}, and R.~{Schöne}, ``{Introducing
  FIRESTARTER: A Processor Stress Test Utility},'' in \emph{2013 International
  Green Computing Conference Proceedings}, 2013,
  \href{https://doi.org/10.1109/IGCC.2013.6604507}{DOI:
  10.1109/IGCC.2013.6604507}.

\bibitem{mprime_website}
\BIBentryALTinterwordspacing
G.~{Woltman}, ``Gimps/mprime website.'' [Online]. Available:
  \url{http://www.mersenne.org/}
\BIBentrySTDinterwordspacing

\bibitem{linpack}
J.~J. Dongarra, P.~Luszczek, and A.~Petitet, ``{The LINPACK Benchmark: past,
  present and future},'' \emph{Concurrency and Computation: Practice and
  Experience}, 2003, \href{https://doi.org/10.1002/cpe.728}{DOI:
  10.1002/cpe.728}.

\bibitem{stressng_website}
\BIBentryALTinterwordspacing
C.~I. King, ``Stress-ng website.'' [Online]. Available:
  \url{https://kernel.ubuntu.com/~cking/stress-ng/}
\BIBentrySTDinterwordspacing

\bibitem{Molka_12_eeMark}
D.~Molka, D.~Hackenberg, R.~Sch\"{o}ne, T.~Minartz, and W.~E. Nagel,
  ``{Flexible Workload Generation for HPC Cluster Efficiency Benchmarking},''
  \emph{Computer Science - Research and Development volume}, 2012,
  \href{https://doi.org/10.1007/s00450-011-0194-9}{DOI:
  10.1007/s00450-011-0194-9}.

\bibitem{Dennard_07_Scaling}
R.~H. Dennard, J.~Cai, and A.~Kumar, ``{A perspective on today’s scaling
  challenges and possible future directions},'' \emph{Solid-State Electronics},
  2007, special Issue: Papers selected from the 2006 ULIS Conference.

\bibitem{Weste_2010_CMOS_VLSI}
N.~Weste and D.~Harris, \emph{{CMOS VLSI Design: A Circuits and Systems
  Perspective}}, 4th~ed.\hskip 1em plus 0.5em minus 0.4em\relax Addison-Wesley
  Publishing Company, 2010.

\bibitem{dvfs}
T.~D. Burd, T.~A. Pering, A.~J. Stratakos, and R.~W. Brodersen, ``{A Dynamic
  Voltage Scaled Microprocessor System},'' \emph{IEEE Journal of Solid-State
  Circuits}, 2000, \href{https://doi.org/10.1109/4.881202}{DOI:
  10.1109/4.881202}.

\bibitem{Molka_2010_Energy}
D.~Molka, D.~Hackenberg, R.~Schöne, and M.~S. Müller, ``{Characterizing the
  Energy Consumption of Data Transfers and Arithmetic Operations on x86-64
  Processors},'' in \emph{International Conference on Green Computing}, 2010,
  \href{https://doi.org/10.1109/GREENCOMP.2010.5598316}{DOI:
  10.1109/GREENCOMP.2010.5598316}.

\bibitem{Lucas_2016_AluPower}
J.~Lucas and B.~Juurlink, ``{ALUPower: Data Dependent Power Consumption in
  GPUs},'' in \emph{2016 IEEE 24th International Symposium on Modeling,
  Analysis and Simulation of Computer and Telecommunication Systems (MASCOTS)},
  2016, \href{https://doi.org/10.1109/MASCOTS.2016.21}{DOI:
  10.1109/MASCOTS.2016.21}.

\bibitem{Schoene_2019_SKL}
R.~Schöne, T.~Ilsche, M.~Bielert, A.~Gocht, and D.~Hackenberg, ``{Energy
  Efficiency Features of the Intel Skylake-SP Processor and Their Impact on
  Performance},'' in \emph{International Conference on High Performance
  Computing Simulation (HPCS)}, 2019,
  \href{https://doi.org/10.1109/HPCS48598.2019.9188239}{DOI:
  10.1109/HPCS48598.2019.9188239}.

\bibitem{Rountree_2021_Performance_Power_Bound}
B.~Rountree, D.~H. Ahn, B.~R. de~Supinski, D.~K. Lowenthal, and M.~Schulz,
  ``{Beyond DVFS: A First Look at Performance under a Hardware-Enforced Power
  Bound},'' in \emph{2012 IEEE 26th International Parallel and Distributed
  Processing Symposium Workshops PhD Forum}, 2012,
  \href{https://doi.org/10.1109/IPDPSW.2012.116}{DOI: 10.1109/IPDPSW.2012.116}.

\bibitem{Hackenberg_2015_Haswell}
D.~Hackenberg, R.~Sch{\"{o}}ne, T.~Ilsche, D.~Molka, J.~Schuchart, and
  R.~Geyer, ``{An Energy Efficiency Feature Survey of the {Intel} {Haswell}
  Processor},'' in \emph{2015 {IEEE} International Parallel and Distributed
  Processing Symposium Workshop}, 2015,
  \href{https://doi.org/10.1109/ipdpsw.2015.70}{DOI: 10.1109/ipdpsw.2015.70}.

\bibitem{stressng_presentation}
\BIBentryALTinterwordspacing
C.~I. King, ``{Stress Testing and Micro Benchmarking Kernels with Stress-ng
  (presentation)},'' 2019. [Online]. Available:
  \url{https://elinux.org/images/5/5c/Lyon-stress-ng-presentation-oct-2019.pdf}
\BIBentrySTDinterwordspacing

\bibitem{Kobalicek_AsmJit}
\BIBentryALTinterwordspacing
P.~Kobalicek, \emph{AsmJit}. [Online]. Available:
  \url{https://github.com/asmjit/asmjit}
\BIBentrySTDinterwordspacing

\bibitem{Hoehlig_2016}
S.~H\"{o}hlig, ``{H{\"o}chstbelastung von Prozessoren auf Basis von automatisch
  generierten Code-Sequenzen},'' Master's thesis, TU Dresden, 2016.

\bibitem{Deb_2002_NSGA-II}
K.~Deb, A.~Pratap, S.~Agarwal, and T.~Meyarivan, ``{A fast and elitist
  multiobjective genetic algorithm: NSGA-II},'' \emph{IEEE Transactions on
  Evolutionary Computation}, 2002,
  \href{https://doi.org/10.1109/4235.996017}{DOI: 10.1109/4235.996017}.

\bibitem{Hackenberg_2013_Power}
D.~Hackenberg, T.~Ilsche, R.~Schöne, D.~Molka, M.~Schmidt, and W.~E. Nagel,
  ``{Power Measurement Techniques on Standard Compute Nodes: A Quantitative
  Comparison},'' in \emph{2013 IEEE International Symposium on Performance
  Analysis of Systems and Software (ISPASS)}, 2013,
  \href{https://doi.org/10.1109/ISPASS.2013.6557170}{DOI:
  10.1109/ISPASS.2013.6557170}.

\bibitem{Schoene_2021_Zen2}
R.~Sch\"{o}ne, T.~Ilsche, M.~Bielert, M.~Velten, M.~Schmidl, and D.~Hackenberg,
  ``{Energy Efficiency Aspects of the AMD Zen 2 Architecture},'' in \emph{IEEE
  International Conference on Cluster Computing (CLUSTER)}.\hskip 1em plus
  0.5em minus 0.4em\relax IEEE, 2021, (accepted).

\bibitem{hickmann_bradford_fletcher_2016}
B.~Hickmann, D.~Bradford, and T.~Fletcher, ``Reducing power consumption in a
  fused multiply-add (fma) unit responsive to input data values,'' U.S. Patent
  9\,323\,500B2, Apr 26, 2016.

\bibitem{Ilsche_2019_MetricQ}
T.~Ilsche, D.~Hackenberg, R.~Sch{\"{o}}ne, M.~Bielert, F.~H{\"{o}}pfner, and
  W.~E. Nagel, ``{MetricQ: A Scalable Infrastructure for Processing
  High-Resolution Time Series Data},'' in \emph{2019 IEEE/ACM
  Industry/University Joint International Workshop on Data-center Automation,
  Analytics, and Control (DAAC)}, 2019,
  \href{https://doi.org/10.1109/DAAC49578.2019.00007}{DOI:
  10.1109/DAAC49578.2019.00007}.

\bibitem{AMD_PPR_Server}
{Advanced Micro Devices}, ``{Preliminary Processor Programming Reference (PPR)
  for AMD Family 17h Model 31h, Revision B0 Processors},'' 2020,
  \url{https://developer.amd.com/wp-content/resources/55803_B0_PUB_0_91.pdf}(accessed
  2021-06-18).

\bibitem{AMD_Zen2_Overview}
D.~{Suggs}, M.~{Subramony}, and D.~{Bouvier}, ``{The AMD “Zen 2”
  Processor},'' \emph{IEEE Micro}, 2020,
  \href{https://doi.org/10.1109/MM.2020.2974217}{DOI: 10.1109/MM.2020.2974217}.

\end{thebibliography}
	
\end{document}